\documentclass[journal=jacsat,manuscript=article]{achemso}

\usepackage[version=3]{mhchem} % Formula subscripts using \ce{}

\author{Bin Cui}
\affiliation{School of Physics, State Key Laboratory of Crystal Materials, Shandong University, Jinan 250100, China}
\alsoaffiliation{Department of Materials Science and Engineering, University of Utah, Salt Lake City, UT 84112, USA}
\author{Bing Huang}
\email{bing.huang@csrc.ac.cn}
\affiliation{Beijing Computational Science Research Center, Beijing 100193, China}
\alsoaffiliation{Department of Materials Science and Engineering, University of Utah, Salt Lake City, UT 84112, USA}
\author{Chong Li}
\affiliation{Department of Materials Science and Engineering, University of Utah, Salt Lake City, UT 84112, USA}
\author{Xiaoming Zhang}
\affiliation{School of Physics, State Key Laboratory of Crystal Materials, Shandong University, Jinan 250100, China}
\alsoaffiliation{Department of Materials Science and Engineering, University of Utah, Salt Lake City, UT 84112, USA}
\author{Kyung-Hwan Jin}
\affiliation{Department of Materials Science and Engineering, University of Utah, Salt Lake City, UT 84112, USA}
\author{Lizhi Zhang}
\affiliation{Department of Materials Science and Engineering, University of Utah, Salt Lake City, UT 84112, USA}
\author{Wei Jiang}
\affiliation{Department of Materials Science and Engineering, University of Utah, Salt Lake City, UT 84112, USA}
\author{Desheng Liu}
\affiliation{School of Physics, State Key Laboratory of Crystal Materials, Shandong University, Jinan 250100, China}
\alsoaffiliation{Department of Physics, Jining University, Qufu 273155, China}
\author{Feng Liu}
\email{fliu@eng.utah.edu}
\affiliation{Department of Materials Science and Engineering, University of Utah, Salt Lake City, UT 84112, USA}
\alsoaffiliation{Collaborative Innovation Center of Quantum Matter, Beijing 100084, China}
\title[An \textsf{achemso} demo]
  {Creation of Half-metallic $f$-orbital Dirac Fermion with Superlight Elements in Orbital-Designed Molecular Lattice}

\abbreviations{C20,TB,SpinFET}

\begin{document}
%%%%%%%%%%%%%%%%%%%%%%%%%%%%%%%%%%%%%%%%%%%%%%%%%%%%%%%%%%%%%%%%%%%%%
%% The manuscript does not need to include \maketitle, which is
%% executed automatically.  The document should begin with an
%% abstract, if appropriate.  If one is given and should not be, the
%% contents will be gobbled.
%%%%%%%%%%%%%%%%%%%%%%%%%%%%%%%%%%%%%%%%%%%%%%%%%%%%%%%%%%%%%%%%%%%%%
\begin{abstract}
  Magnetism in solids generally originates from the localized $d$- or $f$-orbitals that are hosted by heavy transition-metal elements. Here, we demonstrate a novel mechanism for designing half-metallic $f$-orbital Dirac fermion from superlight $sp$-elements. Combining first-principles and model calculations, we show that bare and flat-band-sandwiched (FBS) Dirac bands can be created when C$_{20}$ molecules are deposited into a 2D hexagonal lattice, which are composed of $f$-molecular orbitals (MOs) derived from $sp$-atomic orbitals (AOs). Furthermore, charge doping of the FBS Dirac bands induces spontaneous spin-polarization, converting the system into a half-metallic Dirac state. Based on this discovery, a model of spin field effect transistor is proposed to generate and transport 100\% spin-polarized carriers. Our finding illustrates a novel concept to realize exotic quantum states by manipulating MOs, instead of AOs, in orbital-designed molecular crystal lattices.
\end{abstract}

Magnetic ordering in solids is generally associated with heavy elements, and tremendous efforts have been devoted to finding transition metal (TM) based ferromagnetic (FM) materials\cite{Dietl_RevModPhys2014, Wolf_science2001} for spintronic device applications in the past decades.  It has been known for a long time, according to Stoner criterion\cite{Stoner-1938}, that the origin of FM order in solids arises from the partially occupied (localized) $d$- or $f$-orbitals, which gives rise to a significantly large density of state (DOS) at Fermi level ($E_F$), and drive the system into a FM ground state. Half-metal\cite{DeGroot_prl1983,VanLeuken_prl1995}, with  a 100\% spin-polarized FM state, i.e., conducting in one spin channel and insulating in the other, is an ideal candidate\cite{Jedema_Nature2001} for spin injection and filtering. However, half-metals are very rare and only three-dimensional (3D) Heusler compounds\cite{Dong_APL1999,Jourdan_Nat.Comm.2014}, binary metal oxides CrO$_2$\cite{Watts_prb2000} and mixed valence perovskites\cite{Park_Nature1998} have been identified as half-metals in experiments.

Unfortunately, the strong spin scattering induced by the large spin-orbit coupling (SOC) of heavy elements may significantly reduce the spin relaxation time and mean free path during spin transport. As a result, great attention has been attracted to searching for $sp$-electron magnetism in place of $d$- or $f$- electron magnetism. Local FM order and even half-metallic state have been predicted and confirmed to exist in graphene-related 2D systems, some involving local defects or impurities\cite{1stZGNRafm_jpsj1996,Son2006, BGNR2007prl, Kan_jacs2008, CVacFM_2003prl, FMzgnr_2004NatLett}. However, the realization of local magnetic states in graphene-related systems requires either large external electric fields or well-ordered defects/dopants. It is therefore highly desirable to discover new mechanisms for realizing delocalized FM states in $sp$-electron systems beyond the graphene models.

In this Communication, we demonstrate a novel concept for designing homogenous and long-range magnetism in an $sp$-electron molecular crystal. Our design principle is twofold: first, beyond atomic crystals, molecular crystals are proposed to form with pre-defined lattice symmetries by taking the advantages of chemical synthesis and epitaxial growth\cite{Stock2012,hexC20-2001,fccC20Iqbal-2003,Roy2013,Plas2016,Rochford_jpcl2016}; second, overcoming the fixed orbital energy sequence of atomic orbitals (AOs), molecular orbitals (MOs) having different orbital energy and sequences are designed to have  artificial frontier orbitals with large angular momenta (e.g. $d$ and $f$) in a given crystal field to exhibit exotic quantum features, including magnetism\cite{Tomalia_CR2016}.

Specifically, we illustrate this novel design concept by first-principles and model calculations which show the existence of a half-metallic Dirac semimetal state in 2D molecular crystals made of fullerenes. Taking C$_{20}$ molecules as an example, once these ``superatoms" are arranged into a hexagonal lattice, our first-principles calculations show that the frontier $f$-MOs form bare and flat-band-sandwiched (FBS) Dirac bands around the band gap of the 2D-C$_{20}$ layer. Interestingly, additional fractional charge doping induces a spontaneous spin-polarization of FBS Dirac bands, giving rise to a half-metallic Dirac semimetal state. For potential device applications, the mechanism of our findings can be applied to design a spin field effect transistor (SFET) built from the 2D-C$_{20}$ layer, to generate and transport 100\% spin-polarized carriers.

%fig01
\begin{figure}[tbp]
     \includegraphics[width=15cm]{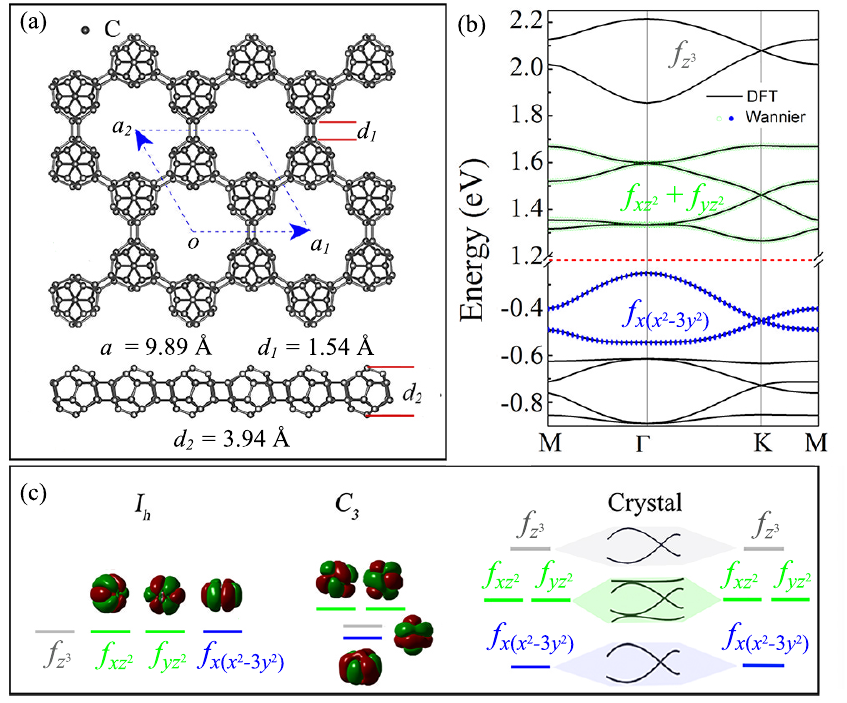}
     \caption{(Color online) (a) Top (upper) and side (bottom) views of C$_{20}$ superatomic honeycomb lattices. The dashed frame indicates the unitcell, along with the marked unit vectors $a_1$ and $a_2$. $d_1$ defines the distance between two neighboring C$_{20}$ molecules and $d_2$ defines the thickness of 2D-C$_{20}$ layer. (b) The DFT calculated band structure of 2D-C$_{20}$ layer. The WLMF-fitted bare and FBS Dirac bands, marked as dashed lines, are also plotted for comparison. The $E_F$ is marked by the red dashed line. (c) Schematic illustration of the mechanism for the formation of bare (FBS) Dirac bands. From left to right: the frontier $f$-MOs of a individual C$_{20}$ molecules with $I_h$ (left) and $C_3$ (middle) symmetries calculated by the Gaussian 09 package, and the bare and FBS Dirac bands from single- and double- $f$-MOs in the 2D-C$_{20}$ crystal (right), respectively.}
      \label{fig_1}
\end{figure}

All the first-principles density functional theory (DFT) calculations are performed using the plane-wave basis Vienna \textit{ab initio} simulation package (VASP)\cite{Kresse_cms1996,Kresse_prb1996} with Perdew-Burke-Ernzerhof exchange correlation functional\cite{PBE_1996}. The cutoff energy for plane wave is chosen to be 400 eV and the vacuum space is set to 15 \AA. All the atoms in the unit cell are relaxed until the forces are smaller than 0.01 eV/\AA. Our test calculations show that SOC has a negligible effect on the electronic properties of C$_{20}$ systems.  The DFT results are used as input to construct the maximally localized Wannier funcitons (MLWFs) with the WANNIER90 code\cite{W90Code}. The MOs of single C$_{20}$ molecule is calculated using Gaussian09 package at b3lyp/6-31g level. The charge conductance of 2D-C$_{20}$ layer are calculated using the DFT non-equilibrium Green's Function method, as implemented in Atomistix ToolKit (ATK) package.

The C$_{20}$ molecule, which was first synthesized in 2000\cite{1stC20Prinzbach-2000}, is known as the smallest fullerene, and it has a dodecahedral cage structure. Important for our purpose, C$_{20}$ molecules can form molecular crystals with various structural symmetries, such as hexagonal and fcc crystals\cite{hexC20-2001, fccC20Iqbal-2003}. When C$_{20}$ molecules are constrained into a 2D lattice, our extensive structural searches based on DFT calculations confirm that the hexagonal lattice is the most stable one, as shown in Figure \ref{fig_1}a. In this structure, two single C-C bonds ($d_1=1.53$ \AA) between the neighboring C$_{20}$ molecules are formed. The calculated equilibrium lattice constant is $a=9.89$ \AA~ and thickness $d_2=3.94$ \AA.

Figure \ref{fig_1}b shows the calculated band structure of 2D-C$_{20}$ lattice. It is a nonmagnetic (NM) semiconductor with an indirect band gap of 1.5 eV. Two subsets of bare Dirac bands and two subsets of FBS Dirac bands appear around the band gap. These Dirac bands appear in different energy ranges without overlapping, and the $E_F$ lies in the gap between two sets of bare and FBS Dirac bands. As we learned from graphene models, the appearance of bare (FBS) Dirac bands could be a result of single $p_z$ (double $p_x+p_y$) orbital hopping on a hexagonal lattice\cite{Wallace1947, Reich2002, Wu2007, prlZLiu-2013}. However, in our C$_{20}$ system all the $sp$-AOs from forty C atoms (per unitcell) contribute to the formation of these Dirac bands and flat bands, as shown in Figure S1\cite{SM}. Thus, it is impossible to understand these Dirac bands by the simple single- or double-AO hopping mechanism developed in graphene models.

Individual C$_{20}$ ``superatom" has a $I_h$ symmetry in its gas phase (C$_{20}^{+2}$), and its MOs are arranged in the order of $1s^22p^63d^{10}4f^{14}5g^{18}6h^{10}2s^23p^64d^{10}5f^2$\cite{JelliumC20_2012} based on Gaussian09 calculations, as shown in Figure S2\cite{SM}, which is significantly different from that of single C atom. Based on electron counting, the four degenerated highest occupied MOs (HOMOs) are of $5f$-type and occupied by two electrons, as shown in Figure 1c (left panel). When C$_{20}$ molecules are arranged into a hexagonal lattice, these $5f$-MOs are splitted into two single $f_{z^3}$ and $f_{x(x^2-3y^2)}$ orbitals plus one double-degenerated $f_{xz^2}+f_{yz^2}$ orbital according to $C_{3}$ crystal field symmetry, as shown in Figure 1c (middle panel). From orbital symmetry analysis, one finds that the $f_{z^3}$ or $f_{x(x^2-3y^2)}$ ($f_{xz^2}+f_{yz^2}$) orbital has a similar in-plane symmetry as the $p_z$($p_x+p_y$) orbital in a hexagonal lattice. It is, therefore, reasonable to expect that hopping between $f_{z^3}$ or $f_{x(x^2-3y^2)}$ ($f_{xz^2}+f_{yz^2}$) orbitals will share similar features as that of $p_z$ ($p_x+p_y$) orbitals. Indeed, our Wannier function calculations show that one (two) MLWF orbital(s) of $f_{z^3}$ or $f_{x(x^2-3y^2)}$ ($f_{xz^2}+f_{yz^2}$), located at the center of the carbon cage, can be uniquely applied to fit perfectly the bare (FBS) Dirac bands around the $E_F$, confirming our physical intuition (right panel of Figure 1c).

%fig02
\begin{figure}[tbp]
     \includegraphics[width=12cm]{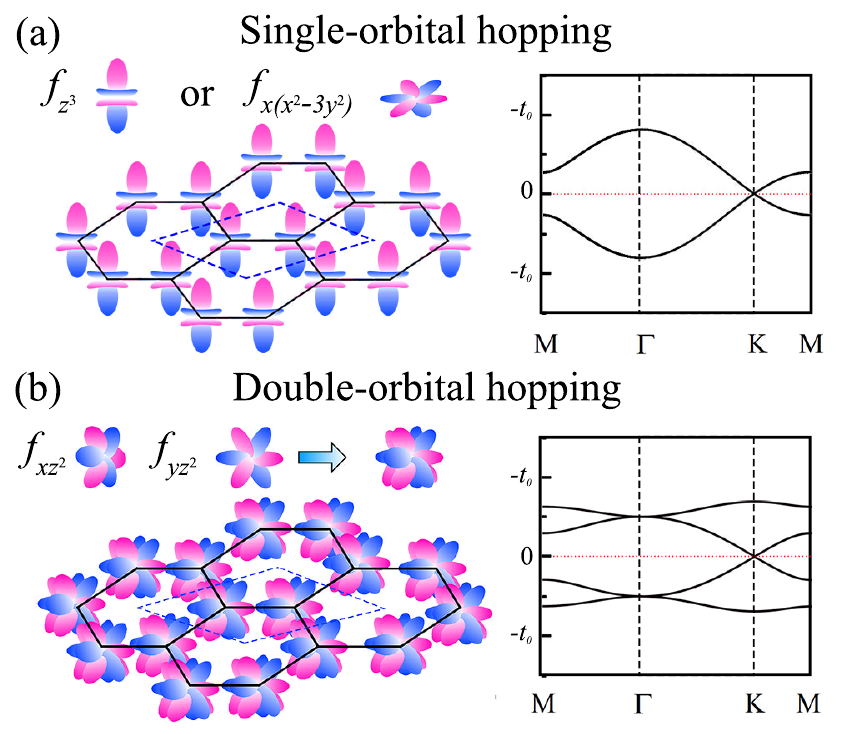}
     \caption{(Color online) \emph{f}-orbital tight-binding Dirac bands: (a) Left panel: schematic picture of $f_{z^3}$ or $f_{x(x^2-3y^2)}$ single-orbital hopping mechanism in a hexagonal lattice. Right panel: the corresponding tight-binding calculated two-band band structure. (b) Left panel: schematic picture of $f_{xz^2}+f_{yz^2}$ double-orbital hopping mechanism in a hexagonal lattice. Right panel: the corresponding tight-binding calculated four-band band structure. All the parameters are described in the text.}
      \label{fig_2}
\end{figure}

In order to have a more general understanding of the formation of bare and FBS $f$-orbital Dirac bands, as discovered in the 2D-C$_{20}$ layer, we further develop a single- and double-MO hopping mechanism based on a nearest neighbor (NN) hopping tight-binding (TB) model, as shown in Figure \ref{fig_2}. Here, the general set of $f$-orbital based Slater-Koster hopping integrals\cite{SKtb1954} are adopted, and a $2\times2$ ($4\times4$) matrix is constructed to describe the two-band (four-band) single- (double-) $f_{z^3} \text{ or } f_{x(x^2-3y^2)}$ ($f_{xz^2}+f_{yz^2}$) orbital hopping in a hexagonal lattice, which can be written as:

\begin{equation}
H=
\begin{pmatrix}
E_{1} & T \\
T^* & E_{2}
\end{pmatrix}
\label{1}
\end{equation}

For simplicity, the on-site energies are set to zero ($E_{1}=E_{2}=0$), and the off-diagonal terms capture the NN hoppings. For the $f_{z^3}$ single-orbital case [$f_{x(x^2-3y^2)}$ is similar to $f_{z^3}$], the hopping integral can be described as $T=\xi(\eta+\xi^{-3}e^{\text{i}ky\pi})[\frac{3}{8}(ff\pi)+\frac{5}{8}(ff\phi)]$. For the $f_{xz^2}+f_{yz^2}$ double-orbital case, the hopping integrals can be described as:
\begin{equation}
T=
\begin{pmatrix}
T_{11} & T_{12} \\
T_{12} & T_{22}
\end{pmatrix}
\label{2}
\end{equation}

where $T_{11} = \frac{1}{64} [6 (4 \xi^{-2} + \xi\eta) (ff\sigma) + 3 \xi\eta (ff\pi) + 10 (4 \xi^{-2} + \xi\eta) (ff\delta) + 45 \xi\eta (ff\phi)]$, $T_{22} =\frac{1}{64} [18 \xi\eta (ff\sigma) + (4 \xi^{-2} + \xi\eta) (ff\pi) + 30 \xi\eta (ff\phi) + 15 (4 \xi^{-2} + \xi\eta) (ff\phi)]$, and $T_{12} = -\frac{\sqrt{3}}{64}  \xi\nu [6 (ff\sigma) - (ff\pi) + 10 (ff\delta) - 15 (ff\phi)]$. Here, $\xi = e^{-i k_x \pi/\sqrt{3}}$, $\eta = e^{i k_y \pi} + e^{-i k_y \pi}$, and $\nu = e^{i k_y \pi} - e^{-i k_y \pi}.$  Because of the \emph{f}-orbital angular momentum having $j=4$, we have four basic hopping terms, i.e., $(ff\sigma)$, $(ff\pi)$, $(ff\delta)$ and $(ff\phi)$. In our calculation, the values of these four hopping terms are set to $(ff\sigma)=1.0$, $(ff\pi)=-0.7$, $(ff\delta)=0.03$, and $(ff\phi)=-0.02$ (in unit of $t_0$), respectively. After the diagonalization of Eq. (1), we obtain two-band and four-band band structures, respectively, as shown in Figs. 2a and 2b (right panel). Our simple TB models can faithfully reproduce both bare and FBS Dirac bands as obtained from DFT calculations (Figure 1b). The main features of bare and FBS Dirac bands are insensitive to the detailed values of hopping parameters (except for the slight variation of band dispersion and widths), indicating that they are very robust and generally exist in many $f$-orbital systems with similar lattice symmetry.

%fig03

\begin{figure}[tbp]
     \includegraphics[width=15cm]{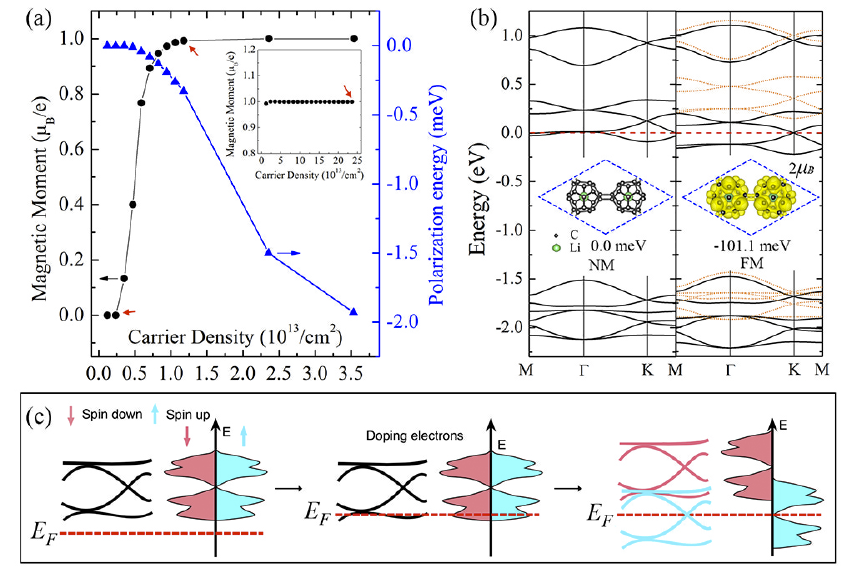}
     \caption{(Color online) (a) The calculated magnetic moment and spin-polarization energy of 2D-C$_{20}$ layer as a function of doped carrier (electron) density. The three red arrows from left to right indicate the critical points for the spin-polarized (left), half-metallic (middle), and half-metallic Dirac semimetal states (right). Inset: the magnetic moment at higher electron concentrations. (b) The calculated band structures of 2D-Li@C$_{20}$ layer for NM (left panel) and FM (right panel) states, respectively. The spin-up and spin-down bands are represented by the (black) solid and (orange) dotted lines, respectively. The $E_F$ is marked by the red dashed line. Inset in left panel: top view of 2D-Li@C$_{20}$ layer; Inset in right panel: the calculated spin density distribution of 2D-Li@C$_{20}$ layer in its FM state. (c) Illustration of the mechanism for the formation of half-metallic semimetal state. From left to right: the semiconducting electronic structure of 2D-C$_{20}$ layer, the partially-filled-flat-band state due to electron doping, and the half-metallic semimetal state mediated by doping of flat bands.}
     \label{fig_3}
\end{figure}

It is important to note that the nontrivial flat band created in 2D-C$_{20}$ layer (wavefunction is delocalized in real space) is significantly different from that created by defects in semiconductors (wavefunction is localized in real space).This unusual flat band at the bottom of the conduction band indicates that a small perturbation of exchange interaction introduced by carrier (electron) doping can induce a large spin splitting, because of the very nature of flat band with an ``infinitly"  large DOS\cite{FBrev_2014}. Accordingly, the critical value of carrier concentration ($n_e$) for spontaneous spin-polarization in 2D-C$_{20}$ system may be much smaller than that in conventional FM materials, according to Stoner criterion. To confirm this expectation, we have calculated the magnetic moment of the system as a function of $n_e$ and the spin-polarization energy (the energy difference between spin-polarized ground state and spin-unpolarized state), which is shown in Figure \ref{fig_3}a. When $n_e$ increases, the flat band becomes partially occupied and the DOS at $E_F$ increases quickly. The critical $n_e$ for spin-polarization in the 2D-C$_{20}$ lattice is estimated to be as small as $\sim2.4\times10^{12}\text{cm}^{-2}$ marked by the left red arrow, which could be achieved in the current experiments by gating (order of $10^{14}\text{cm}^{-2}$)\cite{MoS22011nnano,2015highdoping}. When $n_e>2.4\times10^{12}\text{cm}^{-2}$, the magnetic moment (black circles) continues to increase with the increasing $n_e$, while the spin-polarization energy decreases (blue triangles). Once $n_e>1.18\times10^{13}\text{cm}^{-2}$ marked by the middle red arrow, the magnetic moment saturates at $1.0\mu_B/e$, as shown in the inset of Figure \ref{fig_3}a, and the system becomes a half-metal, and the corresponding band structure is shown in Figure S3\cite{SM}.

Another effective approach to doping of a C$_{20}$ molecule is to encapsulate alkali atoms into the cage of fullerene, which has been achieved in the experiments\cite{Health1985jacs,LiC602010NatChem}. Taking Li as a protoypical example, we insert one Li atom into each C$_{20}$ cage to form a 2D-Li@C$_{20}$ layer, as shown in the inset of Figure \ref{fig_3}b. Overall, the equilibrium geometry and lattice parameters of 2D-C$_{20}$ layer change very little after Li doping. When spin-polarization is excluded (NM case), the calculated band structure is almost the same as the undoped case, except for the shift of $E_F$ to the contacting region between the Dirac bands and the flat band. Including spin-polarization, the half-metallic ground state (right panel of Figure \ref{fig_3}b) is energetically more favorable than the NM state by about 101.1 meV per unitcell. Interestingly, the $E_F$ locates exactly at the Dirac point of the majority (spin-up) channel, creating a half-metallic Dirac semimetal phase with a $\sim0.2$ eV half-metallic gap, and the $n_e$ is  $\sim2.3\times10^{14}\text{cm}^{-2}$, marked by a red arrow in the inset of Figure \ref{fig_3}a. The Fermi velocity of 2D-Li@C$_{20}$ is estimated to be $1.05\times10^5$  m/s, close to that of graphene.

The above findings demonstrate an attractive mechanism for realizing half-metallic Dirac semimetal state by manipulating the MOs of fullerene. First, C$_{20}$ molecule, which has a degenerate $5f$-MOs at HOMO, can be used to generate FBS Dirac bands at the bottom of conduction band under specific crystal field splitting (left panel of Figure 3c). Second, small ``finite" (fractional) charge doping of the flat band will induce a large spin-polarization, because of its ``infinitely" large DOS. As a result, the system is converted from a NM (middle panel of Figure 3c) to a half-metallic ground state (right panel of Figure 3c). Especially, a half-metallic Dirac fermion state can be achieved by a critical $n_e$ by gating or monolayer decoration of alkali atoms.

%fig04
\begin{figure}[tbp]
     \includegraphics[width=12cm]{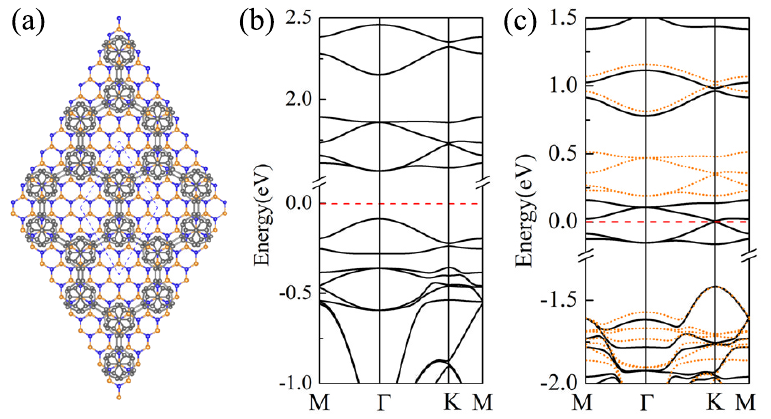}
     \caption{(Color online) (a) Top view of the 2D-C$_{20}$ honeycomb lattice on the monolayer $h$-BN substrate. (b)The calculated band structure of 2D-C$_{20}$ on the $h$-BN substrate. (c) The calculated spin-polarized band structure of 2D-Li@C$_{20}$ on the $h$-BN substrate. The spin-up and spin-down bands are represented by the (black) solid and (orange) dotted lines, respectively. The $E_F$ is marked by the red dashed line.}
      \label{fig_4}
\end{figure}

In practice, the growth of fullerene monolayer with controllable lattice symmetries could be achieved by selection of specific substrates\cite{hexC20-2001,fccC20Iqbal-2003}. Hexagonal BN ($h$-BN) has been widely selected as an ideal substrate to grow various 2D materials by van de Waals (vdW) epitaxy. Using $h$-BN as a candidate substrate, our structure search shows that $1\times1$ 2D-C$_{20}$ (or 2D-Li@C$_{20}$) can perfectly fit a $4\times4$ $h$-BN supercell (lattice mismatch is smaller than 1\%), as shown in Fig. \ref{fig_4}a. Combing the vdW corrections with DFT, the optimized structure of 2D-C$_{20}$  on $h$-BN is shown in Fig. \ref{fig_4}a (the structure of 2D-Li@C$_{20}$ on $h$-BN is similar and not shown here). The minimal distance between C$_{20}$ and the $h$-BN is around 3.2 \AA, indicating a weak interfacial interaction. Our band structure calculations have confirmed that the basic electronic structural characteristics of 2D-C$_{20}$ ($f$-orbtial Dirac bands around bottom of conduction band and top of valence band, as shown in Fig. \ref{fig_4}b) and 2D-Li@C$_{20}$ (half-metallic Dirac band across the Fermi level, as shown in Fig. \ref{fig_4}c) maintain on $h$-BN substrate. Besides $h$-BN, it is expected that 2D-C$_{20}$ could also be grown on other insulating substrates where its exciting electronic structures can be maintained.

%fig05
\begin{figure}[tbp]
     \includegraphics[width=12cm]{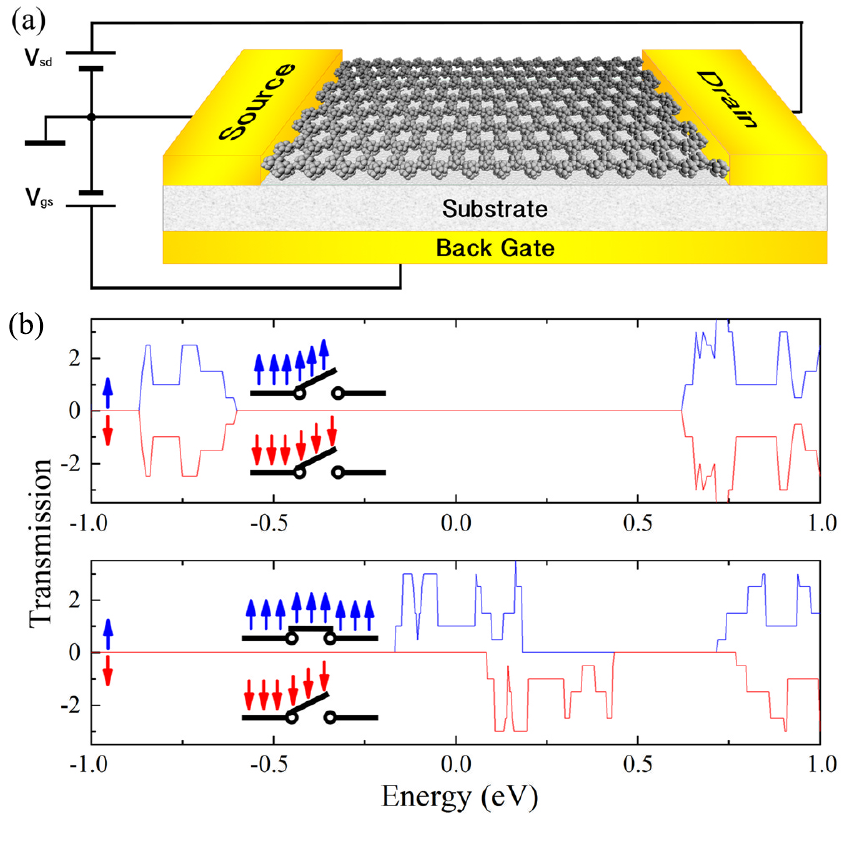}
     \caption{(Color online) (a) The schematic device model of 2D-C$_{20}$ SFET. (b) From top to bottom: the calculated spin-polarized charge conductances for a zero (OFF status) and a nonzero critical (ON status) V$_{gs}$, respectively. The $E_F$ is set to zero.}
      \label{fig_5}
\end{figure}

Once the hexagonal lattice of C$_{20}$ monolayer is grown on a insulating substrate with weak interaction, e.g., the case of C$_{20}$ on $h$-BN, then a 2D-C$_{20}$ based SFET model device can be made, as shown in Figure \ref{fig_5}a. Without gate voltage ($V_{gs}$=0), the 2D-C$_{20}$ monolayer is NM semiconducting with a band gap of $\sim1.5$ eV. Thus, a conductance gap exists around the $E_F$ (OFF status), as shown in Figure \ref{fig_5}b (upper panel). To overcome this transport gap, a bias voltage $V_{sd}\geq1.5$ V has to be applied to drive the spin-unpolarized charge flow through the 2D-C$_{20}$ layer. When the $V_{gs}$ is applied larger than a threshold to achieve the critical doping to occupy the conduction flat band. The system is converted into a half-metallic state (ON status), as shown in Figure \ref{fig_5}b (bottom panel). In this configuration, a small $V_{sd}$ is able to drive a 100\% spin-polarized current through the 2D-C$_{20}$ monolayer. Furthermore, a proper $V_{gs}$ induced charge doping can convert the system into a half-metallic Dirac semimetal state, in which the carriers with 100\% spin-polarization will transport with ultrafast speed.

In conclusion, we demonstrate a novel mechanism for achieving a half-metallic Dirac semimetal state from the manipulation of MOs of fullerenes, by designing large-angular-momentum ($d$- and $f$-) frontier MOs from linear combination of small-angular-momentum ($s$- and $p$-) AOs. Our discovery opens an new avenue to realizing exotic quantum states from orbital-designed molecular crystal lattices.

\begin{acknowledgement}
B.C. acknowledges the support from NSFC and China Scholarship Council (Grant No. 11404188 and 201406225022). B.H., K.J., W.J., and F.L. acknowledge the support from DOE-BES (Grant No. DE-FG02-04ER46148). B.H. also acknowledges the support from NSFC (Grant No.: 11574024) and NSAF U1530401. We thank the CHPC at the University of Utah for providing the computing resources.
\end{acknowledgement}

%\begin{References}

%\end{references}

\begin{suppinfo}

This will usually read something like: ``Experimental procedures and
characterization data for all new compounds. The class will
automatically add a sentence pointing to the information on-line:

\end{suppinfo}

\bibliography{achemso-demo}

\end{document}